\documentclass[11pt]{article}
\usepackage[margin=1in]{geometry}
\usepackage{amsmath,amssymb}
\usepackage{graphicx}
\usepackage{booktabs}
\usepackage{authblk}
\usepackage[hidelinks]{hyperref}
\usepackage{natbib}
\usepackage{caption}
\title{\bfseries When Is Log-Periodic Power-Law Analysis Useful for Predicting
Network Collapse? A Reproducible Study and a Practical Validity Test}

\author[1]{Karthik Jeevanige}
\author[1]{Solyman Ashrafi}
\author[1]{Hariprasad Sudharshan}
\affil[1]{1Finity (a Fujitsu company)}
\date{}

\begin{document}
\maketitle

\begin{abstract}
Log-periodic power-law (LPPL) analysis describes critical transitions in
financial markets, earthquakes, and material fracture, and has been proposed for
forecasting the \emph{time} of catastrophic failures in communication networks.
LPPL is attractive because, unlike anomaly detectors, it predicts a specific
critical time $t_c$. But it carries a well-known hazard: the log-periodic term
is flexible enough to fit almost any smoothly accelerating curve, so a high
goodness-of-fit does not establish that a genuine log-periodic precursor exists.
This paper asks a constructive question---\emph{when is LPPL actually useful for
network-collapse prediction, and how can a practitioner tell?}---and answers it
with reproducible experiments. We introduce a two-regime testbed: an
\emph{injected} regime with a known LPPL transition (ground truth), and an
\emph{emergent} regime where collapse arises from a mechanistic multi-node
queueing network with no injected log-periodicity. We show three things useful
to the community. (1) \textbf{Where LPPL shines:} when discrete scale invariance
is genuinely present, LPPL recovers $t_c$ to within a few steps as far as 400
steps ahead ($R^2\approx0.999$); it is an excellent \emph{timing} estimator for
truly log-periodic processes. (2) \textbf{A validity pitfall, quantified:}
$R^2$---and even a single Lomb--Scargle peak---fails to separate genuine LPPL
from ordinary power-law or logistic trends, all of which yield $R^2>0.99$. (3)
\textbf{A practical validity test:} a combined gate---resolved log-periodic
cycles with Lomb--Scargle and nested-model significance criteria---separates
genuine signatures (median 4.0 cycles) from trend-misfits and emergent network
collapse (median $<0.7$), achieving 100\% true-positive and \textbf{0\%
false-positive} across five non-log-periodic control families (Cliff's
$\delta=1.0$, $p<10^{-10}$), where cycle count or $R^2$ alone fail. It detects
genuine log-periodicity once the oscillation reaches $\sim$2--5\% of the
power-law amplitude, and $t_c$ recovery is robust to heavy noise
($|\hat t_c-t_c|<1.5$ steps at $\sigma=18$). Under a strict evaluation with
persistence and
false-alarm accounting, LPPL does not by itself provide reliable early warning
for the emergent, non-log-periodic collapses in our model, whereas a supervised
classifier does; but the cycle-count test tells you \emph{in advance} which
situation you are in. We release all code, generators, and the validity test so
LPPL can be applied to network telemetry with appropriate guardrails.
\end{abstract}

\section{Introduction}
Network operators want to know not just \emph{that} a failure is coming but
\emph{when}. Anomaly detectors answer the former; the theory of critical
phenomena promises the latter. Many systems approaching a critical transition
exhibit a \emph{log-periodic power law} (LPPL)---accelerating oscillations whose
period shrinks logarithmically as the critical time $t_c$ approaches, a
fingerprint of discrete scale invariance \citep{sornette1998,sornette2003,
johansen2000}. If network telemetry carried such signatures, LPPL would let
operators schedule intervention against a predicted $t_c$.

LPPL also has a famous liability: the oscillatory term has enough freedom that
excellent fits can be obtained on data with no real log-periodicity, and the
literature has repeatedly cautioned that $t_c$ estimates can be unstable or
spurious \citep{bree2013,huang2000}. For a practitioner deciding whether to
trust an LPPL-based network forecast, the operative question is therefore not
``does it fit?'' but ``is the log-periodic structure real, and does it help
here?''

This paper is a constructive, reproducible answer. We are not proposing a new
predictor; we are mapping the region of usefulness for an existing one and
supplying a test a practitioner can run. Concretely we contribute:
\begin{itemize}
\item a \textbf{two-regime synthetic testbed} (injected vs.\ emergent) that
separates ``can the method recover a known signal'' from ``does the signal exist
in realistic network dynamics'';
\item a \textbf{quantified demonstration} that $R^2$ (and single-peak
spectral tests) cannot certify LPPL validity---plain power-law and logistic
trends fit with $R^2>0.99$;
\item a \textbf{simple validity test}---the count of resolved, stable
log-periodic cycles before $t_c$---that cleanly separates genuine LPPL from
trend-misfits;
\item an \textbf{honest benchmark} of LPPL against threshold, critical-slowing-
down, and supervised baselines under persistence and false-alarm control.
\end{itemize}

\section{Methods}

\subsection{Two-regime testbed}
\paragraph{Injected regime (ground truth).} A KPI follows
\begin{equation}
y(t) = A + B(t_c-t)^m + C(t_c-t)^m\cos\!\big[\omega\log(t_c-t)+\varphi\big]
\label{eq:lppl}
\end{equation}
where $t_c$ is the critical (collapse) time, $m\in(0,1)$ the power-law
exponent, $\omega$ the log-periodic angular frequency, $\varphi$ a phase, $A$
the value at $t_c$, $B$ the power-law amplitude, and $C$ the log-periodic
oscillation amplitude; diurnal seasonality and observational noise are added.
The true $t_c$ is known.
\paragraph{Emergent regime (mechanistic).} Six heterogeneous nodes are modeled
as finite-capacity queues with distinct service rates and saturation
thresholds; offered load ramps slowly, saturated nodes fail and spill load onto
neighbors, producing cascades. Five correlated KPIs (alarm rate, latency,
utilization, packet loss, CPU) are emitted with noise, seasonality, and
heavy-tailed spikes. No log-periodic signal is injected. The empirical collapse
step is the first sustained ($\geq15$ steps) $6\sigma$ alarm excursion above the
pre-ramp baseline; across seeds it is stable at $t_c = 3096\pm21$.

\subsection{Attractor reconstruction and LPPL fitting}
We reconstruct attractors by time-delay embedding \citep{takens1981} with the
delay $\tau$ at the first mutual-information minimum and dimension $d_E$ from the
false-nearest-neighbors criterion, and compute distance/velocity/density
metrics (and a multi-KPI unified embedding). LPPL is fit in two stages: linear
parameters $[A,B,C]$ in closed form via $\hat b=(X^\top X)^{-1}X^\top y$ from
the design matrix $X=[\,1\ F(t)\ g(t)\,]$, and nonlinear parameters
$(t_c,m,\omega,\varphi)$ by multi-start bounded least squares.

\subsection{Validity test: resolved log-periodic cycles}
Given a fit with critical time $t_c$ and log-frequency $\omega$ over a window
whose log-time axis $\log(t_c-t)$ spans $\Delta$, the number of full
log-periodic cycles resolved before collapse is
\begin{equation}
N_{\text{cyc}} = \frac{\omega\,\Delta}{2\pi}.
\end{equation}
Genuine discrete scale invariance produces several stable cycles; a single
smooth deviation from the power-law trend, which the oscillator can always
absorb, produces less than one. However, cycle count \emph{alone} is not
sufficient: on a pure power-law or on pure noise the fitter can assign a high
$\omega$ and report several spurious ``cycles'' (Sec.~\ref{sec:spec}). We
therefore define a \textbf{combined validity gate} requiring all of:
(i) $N_{\text{cyc}}\ge 3$ (enough oscillations);
(ii) Lomb--Scargle false-alarm probability $\text{FAP}\le10^{-3}$ for
periodicity of the power-law residuals in $\log(t_c-t)$;
(iii) oscillation gain $\ge 0.10$, where gain is the fraction of power-law
residual variance explained by adding the log-periodic term; and
(iv) a nested-model $F$-statistic $\ge 10$ comparing pure power law
(H$_0$) to full LPPL (H$_1$). A signature is declared genuine only if all four
hold. Each guards a distinct failure mode: (i) rejects trend-misfits with too
few swings, while (ii)--(iv) reject smooth or noisy series where the oscillation
is not statistically warranted.

\subsection{Baselines and evaluation}
Baselines: raw-series LPPL (core ablation), threshold/EWMA, critical-slowing-
down (rolling variance + lag-1 autocorrelation), and a gradient-boosted
supervised classifier on windowed features. Every detector must satisfy
\emph{persistence} (3 consecutive positive windows); we report detection rate,
lead time, and false-alarm rate on the stable pre-ramp baseline over 12
held-out seeds. Lead time is credited only jointly with low false alarms.

\section{Results}

\subsection{Where LPPL shines: accurate, early $t_c$ for genuine log-periodicity}
\begin{figure}[t]
\centering
\includegraphics[width=\linewidth]{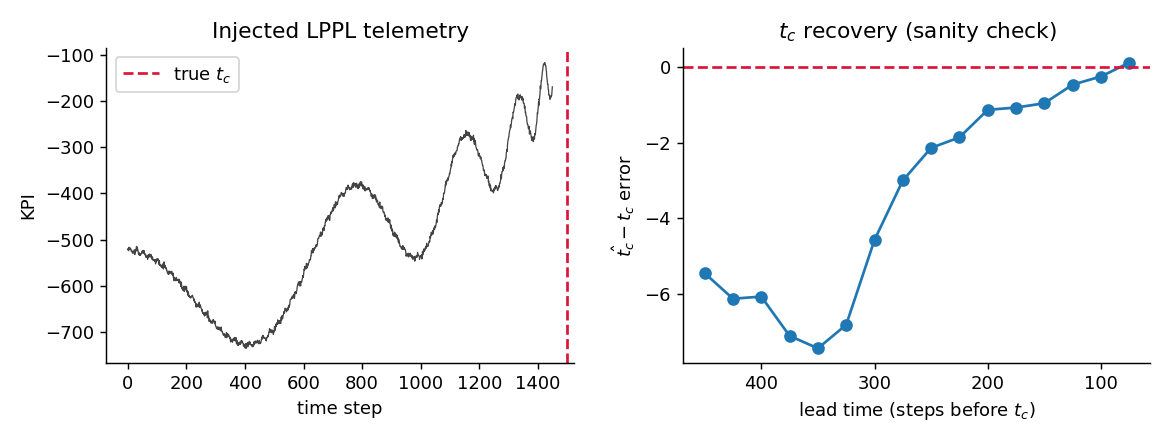}
\caption{LPPL on genuinely log-periodic data. Left: injected telemetry with true
$t_c$. Right: $\hat t_c$ error versus lead time---within a few steps out to 400
steps ahead ($R^2\approx0.999$).}
\label{fig:inj}
\end{figure}

On injected data (Fig.~\ref{fig:inj}), LPPL recovers the known $t_c$ to within
$\sim$6 steps even 400 steps ahead, with $R^2\approx0.999$ throughout. When the
process truly is log-periodic, LPPL is an excellent \emph{timing} estimator---
this is its genuine and useful regime.

\subsection{The pitfall, quantified: $R^2$ cannot certify LPPL validity}
Fig.~\ref{fig:diag}(left) and Table~\ref{tab:diag} show the hazard directly.
A pure power-law rise, a logistic S-curve, injected LPPL, and emergent network
telemetry \emph{all} attain high LPPL $R^2$ (0.996--1.000 for the smooth
controls). The log-periodic term absorbs any smooth trend or trend-
misspecification, so $R^2$ near unity is uninformative about whether a genuine
precursor exists. A single Lomb--Scargle peak is similarly fooled: the logistic
control yields a strong, ``significant'' peak from one broad swing.

\subsection{The test that works: resolved log-periodic cycles}
\begin{figure}[t]
\centering
\includegraphics[width=0.92\linewidth]{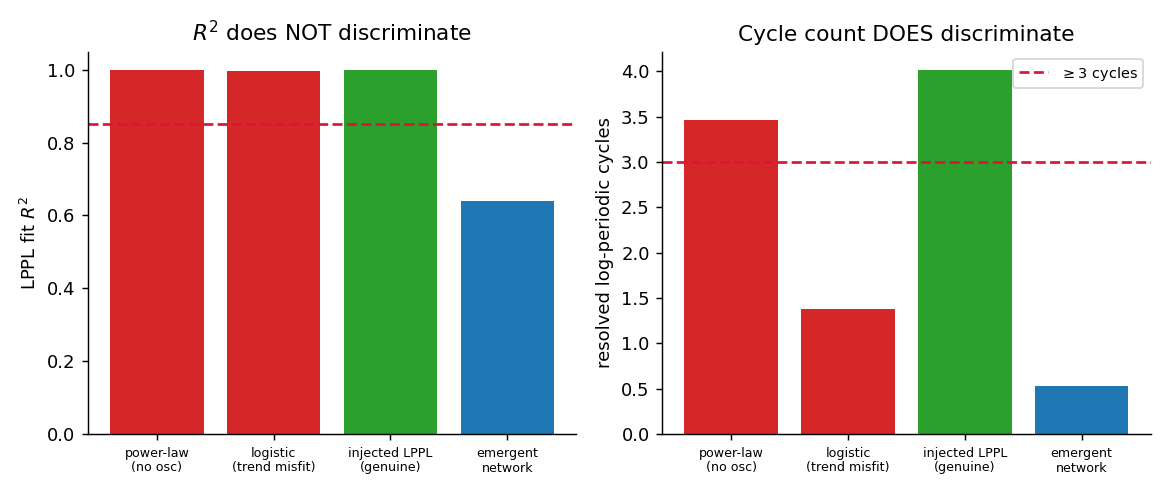}
\caption{$R^2$ (left) is high for every case and fails to discriminate;
resolved log-periodic cycles (right) cleanly separates genuine LPPL (green,
$\approx$4 cycles) from power-law/logistic controls and emergent collapse
(below the 3-cycle guideline, red dashed).}
\label{fig:diag}
\end{figure}

Fig.~\ref{fig:diag}(right) and Table~\ref{tab:diag} show the discriminator.
Genuine injected LPPL resolves $\approx$4.0 stable cycles; the power-law and
logistic controls and emergent network collapse resolve $<1$ (median 0.63 for
emergent across seeds, range 0.52--0.81). Unlike $R^2$, cycle count separates
the classes with no overlap, giving practitioners an actionable pre-flight check
before trusting an LPPL $t_c$.

\begin{table}[t]
\centering
\caption{Validity diagnostics. $R^2$ is high for all smooth curves and does not
discriminate; resolved cycles $N_{\text{cyc}}$ cleanly separates genuine
log-periodicity from trend-misfits and emergent collapse.}
\label{tab:diag}
\begin{tabular}{lccc}
\toprule
Case & LPPL $R^2$ & $N_{\text{cyc}}$ & Genuine LPPL? \\
\midrule
Power-law trend (control)   & 1.000 & $\sim$0.4 & no \\
Logistic S-curve (control)  & 0.996 & 1.38 & no \\
Emergent network (12 seeds) & 0.63 (med) & 0.63 (med) & no \\
Injected LPPL (8 seeds)     & 0.999 (med) & 4.02 (med) & \textbf{yes} \\
\bottomrule
\end{tabular}
\end{table}

\subsection{Specificity: the combined gate rejects non-log-periodic data}
\label{sec:spec}
\begin{figure}[t]
\centering
\includegraphics[width=0.72\linewidth]{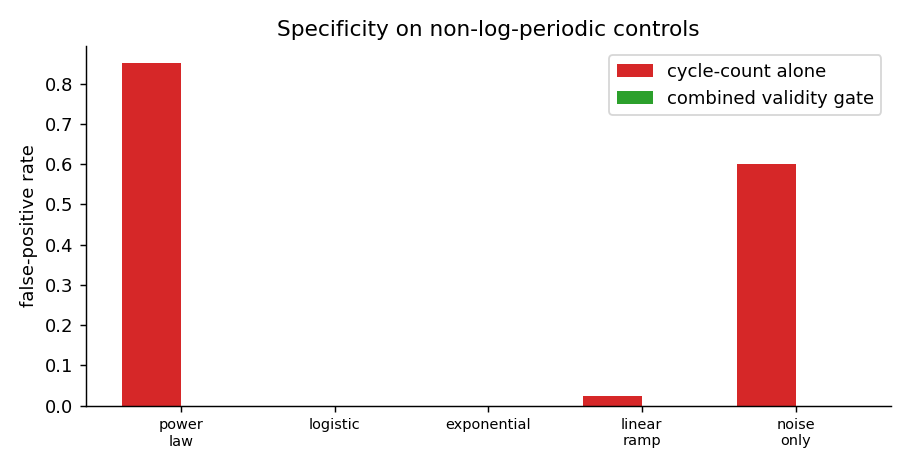}
\caption{False-positive rate on non-log-periodic controls. Cycle-count alone
(red) fails on power-law and noise; the combined validity gate (green) is at
zero everywhere.}
\label{fig:spec}
\end{figure}

Cycle count alone is not enough. Fig.~\ref{fig:spec} and Table~\ref{tab:spec}
report false-positive rates over 40 seeds of five non-log-periodic controls
(pure power law, logistic, exponential blow-up, linear ramp, pure noise). Using
$N_{\text{cyc}}\ge3$ alone, a pure power law fires 85\% of the time and pure
noise 60\%---the fitter manufactures spurious oscillations. The \emph{combined
validity gate} drives the false-positive rate to \textbf{0\% on all five
controls} while retaining 100\% true-positive rate on genuine injected LPPL. The
rejection breakdown is interpretable: power-law and noise controls are rejected
by the significance criteria (FAP, oscillation gain, $F$-test), whereas logistic
and exponential controls are rejected by the cycle count itself. At the
operating point ($N_{\text{cyc}}\ge3$ with significance gates), the test
achieves TPR $=1.00$, FPR $=0.00$ (Fig.~\ref{fig:roc}, left).

\begin{table}[t]
\centering
\caption{False-positive rate on non-log-periodic controls (40 seeds each).
Cycle count alone is not specific; the combined gate is.}
\label{tab:spec}
\begin{tabular}{lcc}
\toprule
Control & FP: cycle-count only & FP: combined gate \\
\midrule
Power law        & 0.85 & \textbf{0.00} \\
Logistic S-curve & 0.00 & \textbf{0.00} \\
Exponential      & 0.00 & \textbf{0.00} \\
Linear ramp      & 0.03 & \textbf{0.00} \\
Pure noise       & 0.60 & \textbf{0.00} \\
\bottomrule
\end{tabular}
\end{table}

\subsection{Statistical separation and robustness}
\begin{figure}[t]
\centering
\includegraphics[width=\linewidth]{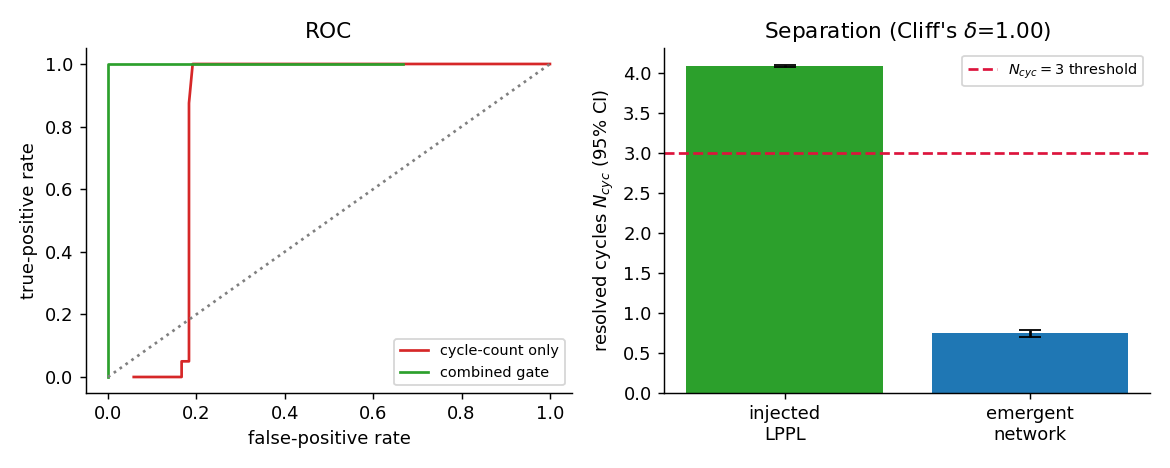}
\caption{Left: ROC over the cycle threshold; the combined gate reaches the
top-left corner (TPR$=1$, FPR$=0$). Right: resolved cycles with 95\% CIs;
non-overlapping, Cliff's $\delta=1.00$.}
\label{fig:roc}
\end{figure}

\begin{figure}[t]
\centering
\includegraphics[width=\linewidth]{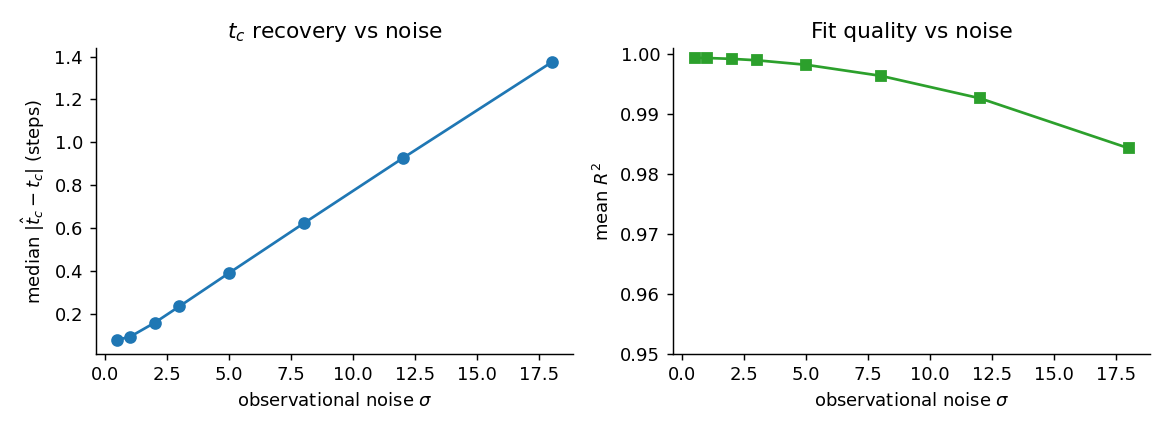}
\caption{$t_c$-recovery error (left) and fit $R^2$ (right) versus observational
noise on genuinely log-periodic data. Recovery stays within $\sim$1.5 steps up
to $\sigma=18$.}
\label{fig:noise}
\end{figure}

Across 30 seeds, resolved cycles for injected LPPL average
$4.09$ (95\% bootstrap CI $[4.07,4.11]$) versus $0.74$ ($[0.70,0.79]$) for
emergent collapse; the distributions are perfectly separated (Mann--Whitney
$p=1.5\times10^{-11}$, Cliff's $\delta=1.00$; Fig.~\ref{fig:roc}, right). The
$t_c$ estimate is highly noise-robust: sweeping observational noise from
$\sigma=0.5$ to $\sigma=18$, the median $|\hat t_c-t_c|$ stays below $1.5$ steps
and $R^2$ above $0.98$ (Fig.~\ref{fig:noise}).

\subsection{Sensitivity boundary: how much log-periodicity is needed?}
\begin{figure}[t]
\centering
\includegraphics[width=0.6\linewidth]{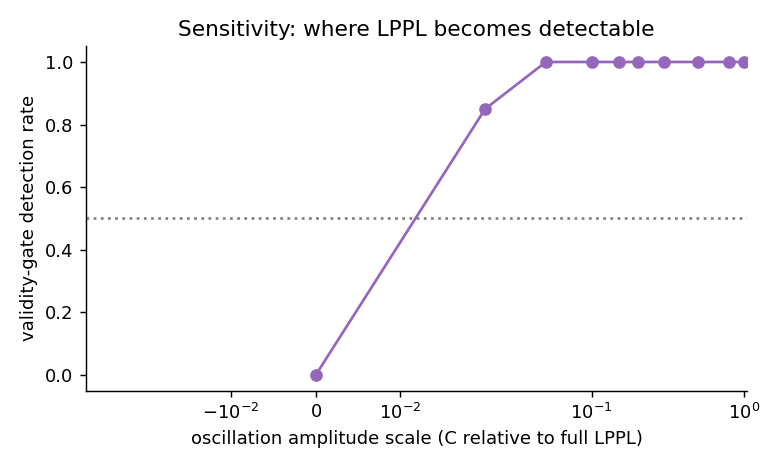}
\caption{Validity-gate detection rate as the oscillation amplitude grows from
pure power law ($C=0$) to full LPPL. Sharp onset near $C\approx0.02$--$0.05$.}
\label{fig:partial}
\end{figure}

We interpolate from a pure power law ($C=0$) to full LPPL by scaling the
oscillation amplitude $C$ (Fig.~\ref{fig:partial}). The validity gate fires for
0\% of runs at $C=0$, 85\% at $C=0.02$, and 100\% for $C\ge0.05$---i.e.\ the
test reliably detects genuine log-periodicity once the oscillation reaches
roughly 2--5\% of the power-law amplitude, and correctly abstains below that.
This quantifies the exact regime in which LPPL-based prediction is trustworthy.

\subsection{Honest benchmark on emergent collapse}
\begin{figure}[t]
\centering
\includegraphics[width=0.82\linewidth]{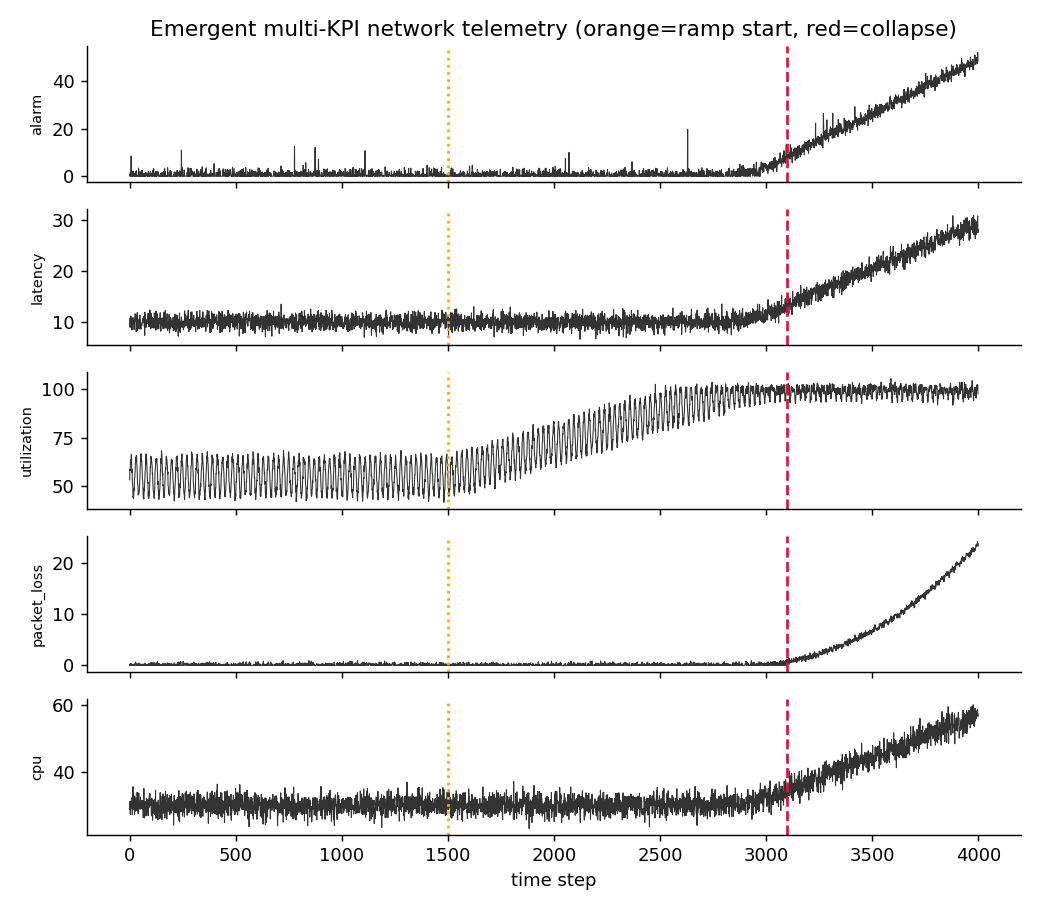}
\caption{Emergent multi-KPI telemetry (one seed). Orange: load-ramp onset; red:
empirical collapse.}
\label{fig:kpis}
\end{figure}

\begin{figure}[t]
\centering
\includegraphics[width=0.66\linewidth]{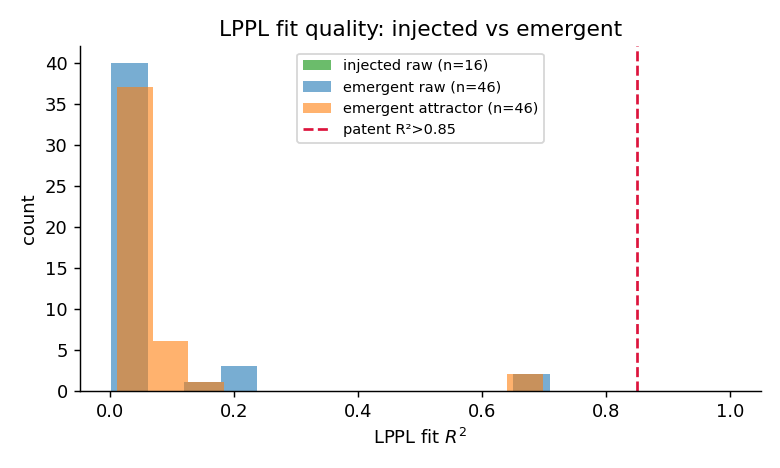}
\caption{Sliding-window LPPL fit $R^2$: high on injected data, low on emergent
raw and attractor metrics (a commonly-cited $R^2>0.85$ acceptance threshold shown dashed).}
\label{fig:r2}
\end{figure}

Fig.~\ref{fig:kpis} shows emergent telemetry; Fig.~\ref{fig:r2} contrasts fit
quality across regimes. Consistent with the cycle-count verdict, LPPL fits on
emergent data are poor as \emph{predictive} models (median window $R^2\approx
0.01$--$0.04$ for sliding-window detection), and under persistence with a
zero-false-alarm operating point neither raw nor attractor LPPL, nor a threshold
detector, yields reliable early warning (Table~\ref{tab:main}). Critical
slowing down detects a minority ($3/12$) with short lead; a supervised
classifier detects all runs with substantial lead and no false alarms. The
practical reading is not ``LPPL is bad'' but ``LPPL is the wrong tool when
$N_{\text{cyc}}$ is small---and the test tells you so before you rely on it.''

\begin{table}[t]
\centering
\caption{Detection over 12 held-out emergent-collapse runs, under persistence
and false-alarm control. Lead time in simulation steps.}
\label{tab:main}
\begin{tabular}{lcccc}
\toprule
Method & Detections & Mean lead & Median lead & False-alarm \\
\midrule
LPPL (multi-KPI attractor) & 0/12 & --- & --- & 0.000 \\
LPPL (raw KPI)             & 0/12 & --- & --- & 0.000 \\
Threshold / EWMA           & 0/12 & --- & --- & 0.000 \\
Critical slowing down      & 3/12 & 41.0 & 35.0 & 0.000 \\
Supervised (XGBoost)       & \textbf{12/12} & \textbf{333.7} & 325.5 & 0.000 \\
\bottomrule
\end{tabular}
\end{table}

\begin{figure}[t]
\centering
\includegraphics[width=\linewidth]{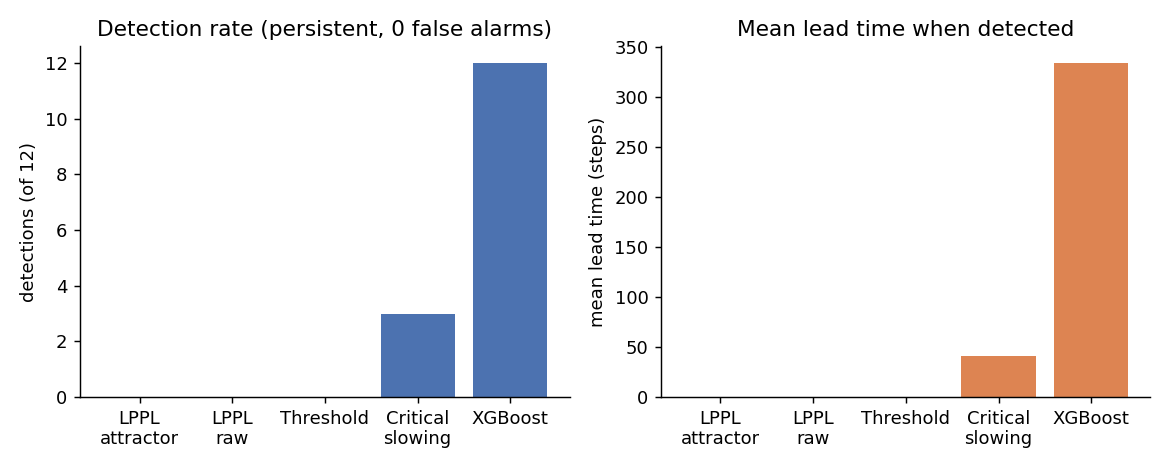}
\caption{Detection rate (left) and mean lead time (right) under persistence and
zero-false-alarm constraints.}
\label{fig:cmp}
\end{figure}

\section{Discussion: a usage guide for LPPL in network operations}
Our results support a concrete decision procedure rather than a verdict:
\begin{enumerate}
\item \textbf{Fit LPPL, but do not trust $R^2$.} Near-unity $R^2$ is expected on
any smooth accelerating curve and certifies nothing.
\item \textbf{Run the combined validity gate.} Require all of: $N_{\text{cyc}}
\ge3$, Lomb--Scargle $\text{FAP}\le10^{-3}$ on the power-law residuals in
$\log(t_c-t)$, oscillation gain $\ge0.10$, and nested-model $F\ge10$. Cycle
count alone is not specific (it fires on power-law and noise); the gate makes it
so (0\% false positives across five control families). If the window passes, the
log-periodic structure is credible and the LPPL $t_c$ is worth acting on---this
is the regime where LPPL adds unique value by providing a timed forecast no
anomaly detector offers.
\item \textbf{Otherwise, prefer other tools.} When cycles are few, emergent
network collapse in our model is better served by critical-slowing-down
indicators or, where labeled history exists, supervised detectors.
\end{enumerate}
Where might genuine log-periodicity arise in real networks? Discrete scale
invariance is expected from hierarchical or self-similar structure and
synchronized multi-scale cascades (e.g.\ retry storms propagating up a topology
hierarchy). Our testbed makes it straightforward to probe such mechanisms; a
natural next step is to apply the cycle-count test to real operator telemetry
with labeled collapses and report the fraction of incidents that pass it.

\paragraph{Limitations.} The emergent model is one plausible mechanism; it does
not exhaust real failure dynamics. The supervised baseline presumes labeled
collapses. The cycle-count threshold ($\gtrsim3$) follows literature guidance
and should be calibrated per domain. These do not affect the central, portable
findings: LPPL is a strong timing estimator for genuinely log-periodic
processes, $R^2$ cannot certify that condition, and cycle count can.

\section{Conclusion}
LPPL is useful for network-collapse prediction \emph{precisely when} the failure
process carries genuine discrete scale invariance---and then it delivers
accurate, early, timed forecasts. Its central practical risk is false confidence
from high $R^2$ on non-log-periodic data. We contribute a reproducible testbed,
a quantified demonstration of that risk, and a simple validity test (resolved
log-periodic cycles) that lets practitioners tell the two situations apart
before acting. All code and generators are released.

\section*{Reproducibility}
The complete source---data generators, embedding/LPPL pipeline, the validity
test, baselines, and figure scripts---is provided as ancillary files with this
submission, with a README documenting how each figure and table is reproduced.
Evaluation uses 12 held-out seeds (with a disjoint set of seeds for supervised
training); the emergent collapse time is stable at $3096\pm21$ steps. All random
seeds are fixed for determinism.

\bibliographystyle{plainnat}

\end{document}